\documentclass{article}
\usepackage[utf8]{inputenc}
\usepackage{graphicx}
\usepackage{xcolor}
\usepackage{geometry}
\usepackage{tabularx}
\usepackage{blindtext}
\usepackage{caption}
\usepackage{float}
\usepackage{url}
\title{Feedback from the participants of the \\ADBIS, TPDL and EDA 2020 joint conferences}
\author{Pegdwendé N. Sawadogo \and Jérôme Darmont \and Fabien Duchateau}
\date{\url{http://eric.univ-lyon2.fr/adbis-tpdl-eda-2020/}}

\begin{document}

\maketitle

\section{Introduction}

The 24\textsuperscript{th} European Conference on Advances in Databases and Information Systems (ADBIS 2020), the 24\textsuperscript{th} International Conference on Theory and Practice of Digital Libraries (TPDL 2020) and the 16\textsuperscript{th} EDA days on Business Intelligence \& Big Data (EDA 2020) were jointly organized from August 25 to August 27, 2020. 
The event also hosted a common Doctoral Consortium (DC) and six workshops: the 1\textsuperscript{st} Workshop on Assessing Impact and Merit in Science (AIMinScience 2020), the 2\textsuperscript{nd} International Workshop on BI \& Big Data Applications (BBIGAP 2020), the International Workshop on Intelligent Data -- From Data to Knowledge (DOING 2020),  Modern Approaches in Data Engineering and Information System Design (MADEISD 2020), the 10\textsuperscript{th} International Symposium on Data-Driven Process Discovery and Analysis (SIMPDA 2020) and Scientific Knowledge Graphs (SKG 2020).

Because of the worldwide pandemic context, the organization committee and respective steering committees decided to organize the event online. 
This choice led to great challenges in choosing the right options and tools to ensure that participants get the most out of the conferences.
We chose to have the conferences take place on about the same days that we had planned for the physical meeting in Lyon. Conference days and talks were shortened, as attending streaming presentations is tiring. The schedule was centered around lunchtime to allow as many participants to attend, especially researchers from North and South America and Asia. Registration fees were decreased accordingly and participants without a paper could register for free.

The videoconferencing system we used was Big Blue Button (BBB)\footnote{\url{https://bigbluebutton.org}}. The platform was hosted at the University of Lyon~2, which allowed us to have more control on both the infrastructure and the system. We asked speakers to prepare a prerecorded presentation in MP4 format with respect to constraints much inspired from the presenter guidelines of the 14\textsuperscript{th} International Baltic Conference on Databases and Information Systems (DB\&IS 2020)\footnote{\url{https://dbis.ttu.ee/index.php?page=53}}. The length of presentations were: keynotes 20 minutes, long papers 10 minutes and short papers 5 minutes. We assisted chairpersons with ``projectionists'' who were in charge of playing the prerecorded presentations. Questions and Answers (Q\&A) sessions after presentations were live, by chat only, and moderated by the chairperson as usual. The duration of Q\&A sessions were: keynotes 10 minutes, long papers  5 minutes and short papers   5 minutes.

Overall, the event appealed to more than 500 registered participants. Yet, attendance was most probably not this high, with a little bit more than 100 maximum connections at the highest peak. We also noticed that the free non-speaker fee attracted ``ghost participants'' who never connected. Hence, setting up even a low registration fee would presumably be a better idea. Detailed attendance figures are provided in Tables~\ref{tab:attendance-workshops}, \ref{tab:attendance-keynotes}, \ref{tab:attendance-sessions} and \ref{tab:attendance-events}.

\begin{table}[hbt]
    \centering
    \begin{tabular}{|c|c|c|c|c|c|c||c|c|}
    \hline
    DC & AIMinScience & BBIGAP & DOING & MADEISD & SIMPDA & SKG & Total & Average \\
    \hline
    37 & 54 & 32 & 46 & 23 & 16 & 64 & 272 & 38.9 \\
    \hline
    \end{tabular}
    \caption{Attendance to the DC and workshops}
    \label{tab:attendance-workshops}
\end{table}

\begin{table}[hbt]
    \centering
    \begin{tabular}{|c|c|c|c|c||c|c|}
    \hline
    Keynote 1 & Keynote 2 & Keynote 3 & Keynote 4 & Keynote 5 & Total & Average \\
    \hline
    63 & 36 & 67 & 66 & 66 & 299 & 59.8\\
    \hline
    \end{tabular}
    \caption{Attendance to keynote speeches}
    \label{tab:attendance-keynotes}
\end{table}

\begin{table}[hbt]
    \centering
    \begin{tabular}{|c|c|c|c|c|c|c|c|c||c|c|}
    \hline
    ADBIS1 & ADBIS2 & ADBIS3 & ADBIS4 & ADBIS5 & TPDL1 & TPDL2 & TPDL3 & EDA & Total & Average \\
    \hline
    40 & 40 & 36 & 38 & 40 & 61 & 44 & 49 & 42 & 390 & 43.3\\
    \hline
    \end{tabular}
    \caption{Attendance to regular sessions}
    \label{tab:attendance-sessions}
\end{table}

\begin{table}[hbt]
    \centering
    \begin{tabular}{|c|c|c|c||c|c|}
    \hline
    Opening session & Virtual reception & Virtual coffee room (1 day) & Closing session & Total & Average \\
    \hline
    57 & 27 & 5 & 55 & 144 & 36.0\\
    \hline
    \end{tabular}
    \caption{Attendance to other events}
    \label{tab:attendance-events}
\end{table}

In order to evaluate what has worked and what has not, we proposed a survey to get the attendants' feedback. We also hope this report can guide the choices for the organization of future conferences. 
The survey whose results are presented thereafter was answered by 106 participants, which amounts to 20.9\% of registered attendants. 
Thus, we believe that the results obtained on the basis of this sample can be considered representative.

\section{Survey Methodology}
\label{sec:methodo}

%\textcolor{red}{Moyen retenu pour le sondage, questions}
The questions we defined in our survey are inspired from the experience of the 36\textsuperscript{th} IEEE International Conference on Data Engineering (ICDE 2020) \cite{bonifati2020} and the 42\textsuperscript{nd} European Conference on Information Retrieval (ECIR 2020) Workshops \cite{nunes2020}. More specifically, survey questions, which explicitly appear in caption of Section~\ref{sec:results}'s figures, can be categorized in five groups.
\begin{enumerate}
    \item \textbf{Questions on participant characteristics.} 
    These questions aim at better knowing the profiles of participants who attended the conferences; and also whether virtualization made easier for non-Europeans to participate. 
    
    \item \textbf{Questions on session attendance.}
    In this category, questions aim at measuring and comparing session audiences, as perceived by participants (we also recorded \textit{in situ} that actual number of participants in each session). This is especially important to identify the attractiveness of sessions and provide hints for next editions.  
    
    \item \textbf{Questions on participant experience.}
    Here, we focus on the participants' appreciation of the conferences' organization. Our goal was to evaluate whether we succeeded in meeting the double challenge of \textit{virtually} organizing \textit{three} joint conferences.
    
    \item \textbf{Questions on participant preferences for future editions.} 
    Since we had to make certain choices when we switched from a live to an online event, we asked \textit{a posteriori} whether participants shared our arbitration. Moreover, we asked participant to suggest better options for future editions in case they were not fully satisfied. 
    
    \item \textbf{Questions on virtualization.}
    We  also introduced the question of whether virtualization could be desirable beyond this pandemic context, notably to reduce the cost of conferences in terms of time, funding and CO2 emissions \cite{OngMS14}.
\end{enumerate}

Eventually, we added a free textual field to collect general comments on the conferences.

The resulting questionnaire has been sent to all participants, including authors/speakers, by email. Let us notice that all responses were anonymous, so no personal information has been collected.

\section{Survey Results}
\label{sec:results}

In this section, we present the findings from the survey through a set of fifteen facts, in response to the five groups of questions from Section~\ref{sec:methodo}.%each of which is detailed and illustrated with figures. 

\subsection{Participant Characteristics}

\begin{figure}[H]
    \centering
    \captionsetup{justification=centering,margin=0cm}
\begin{minipage}{.5\textwidth}
\includegraphics[width=7.5cm]{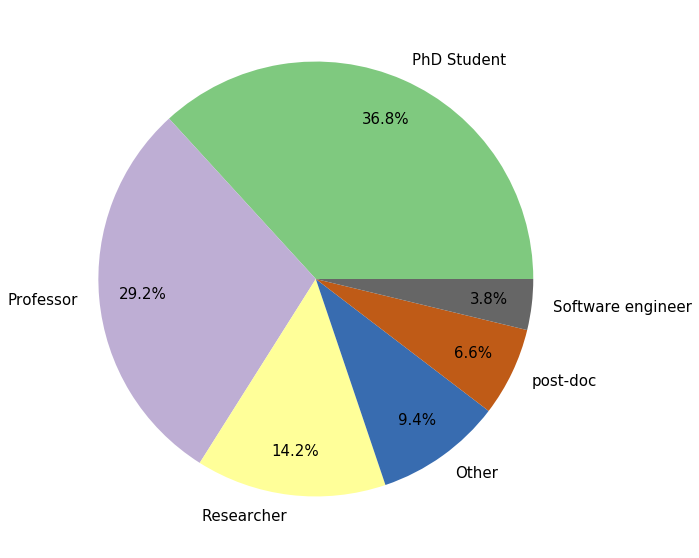} 
\caption{What is your current occupation? \\ (106 responses)}
\label{fig:QA1}
\end{minipage}% This must go next to `\end{minipage}`
\begin{minipage}{.5\textwidth}
\centering
\includegraphics[width=6cm]{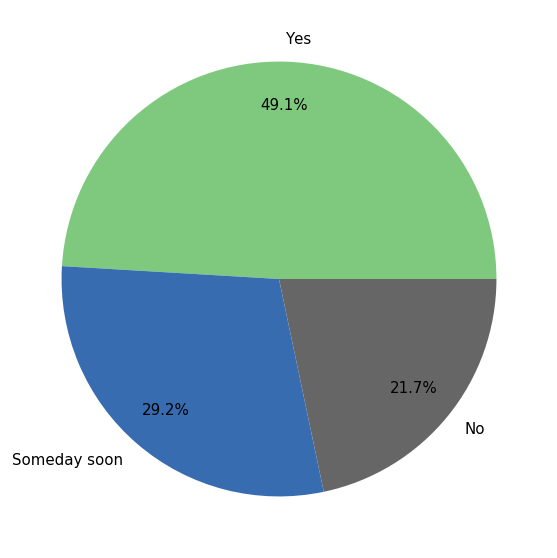} 
\caption{Do you have a PhD? \\ (106 responses)}
\label{fig:QA2}
\end{minipage}
\end{figure}

\paragraph{(1) ADBIS, TPDL and EDA 2020 mostly attracted academic attendants.}
The conferences mainly attracted PhD students (36.8\%) and university professors (29.2\%), as illustrated in Figure~\ref{fig:QA1}. %That shows that ADBIS, TPDL and EDA are globally perceived as mostly dedicated to theoretical than applicative research. 
PhD student attendance is quite high compared to physical conferences (possibily due to early PhD, rejected paper or supervisor traveling to the conference). Hence, it seems that virtualization offers a good and cheap opportunity for PhD student to attend conferences.
Moreover, 78.3\% of attendants were either doctors or about to achieve a PhD (Figure~\ref{fig:QA2}). Yet, considering that software engineers and ``others'' in Figure~\ref{fig:QA1} amount to about 13\% of the population, it leaves 10\% of professors and researchers without a PhD. Although some PhD students may have answered ``No'' instead of ``Someday soon'' by mistake, this figure is surprisingly high.

\begin{figure}[H]
    \centering
    \captionsetup{justification=centering,margin=0cm}
\begin{minipage}{.5\textwidth}
\includegraphics[width=6.5cm]{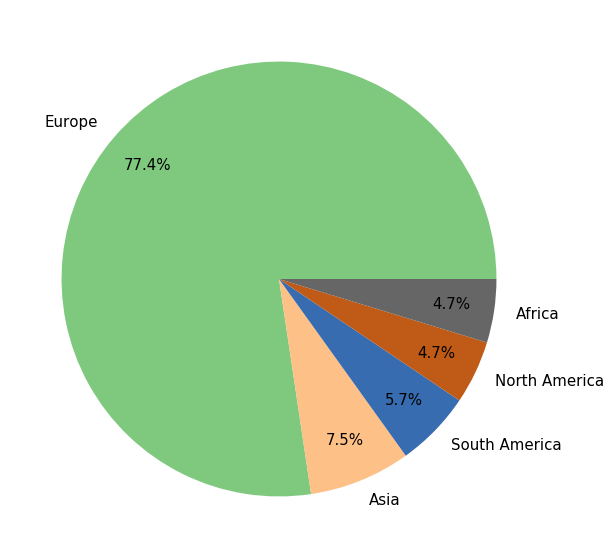} 
\vspace{-0.5cm}
\caption{On what continent were you during \\ the conference? (106 responses)}
\label{fig:QA3}
\end{minipage}% This must go next to `\end{minipage}`
\begin{minipage}{.5\textwidth}
\centering
\includegraphics[width=9cm]{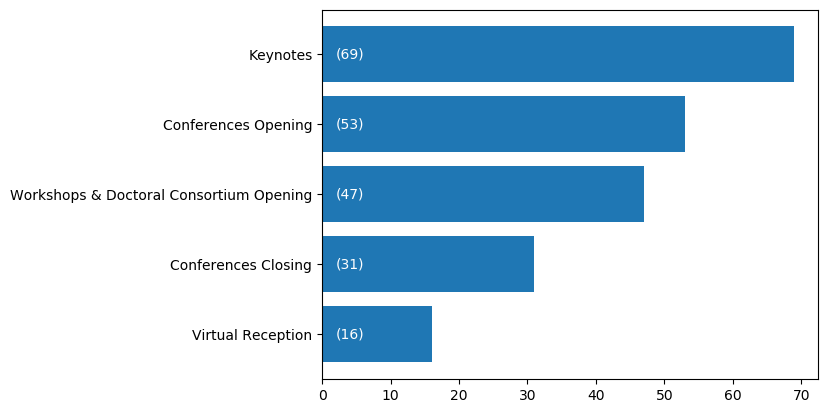} 
\vspace{0.5cm}
\caption{To what plenary sessions did you \\ attend?  (95 responses)}
\label{fig:QB1}
\end{minipage}
\end{figure}

\paragraph{(2) 77.4\% of attendants came from Europe.} ADBIS, TPDL and EDA 2020 reached a majority of European researchers~(Figure~\ref{fig:QA3}), presumably because ADBIS (which is explicitly a European conference) and  TPDL are perceived as European conferences. However, we could have expected a wider diversity of participants with virtualization. This was not the case, probably because the choice to go online was made after paper submissions. The timezone issue may also have had an impact, with participants from Asia and the Americas being able  to attend only a few sessions. Actually, participants who attended to few sessions for any reason may not have wanted to take the survey.

\subsection{Session Attendance}

First, let us remind that this survey was answered by 106 participants, which amounts to 20.9\% of registered attendants. 

\paragraph{(3) Keynotes were very attractive.}
69 out of 95 respondents (72.6\%) attended to ADBIS, TDPL and EDA keynotes (Figure~\ref{fig:QB1}), which confirms what we observed during online sessions, all keynotes but one reaching over 65 attendants. 
This shows keynote speakers were attractive and that their talks interested the three ADBIS, TPDL and EDA communities.   

\paragraph{(4) Satellite events were of particular importance.}
We note that the workshops and doctoral consortium opening attracted almost as many attendants as the  conferences opening (Figure~\ref{fig:QB1}). We can thus consider that such satellite events were important aside the main conferences. 
It is therefore desirable that they are perpetuated in future editions.

\paragraph{(5) ADBIS was the most attended conference.}
With 61 out of 95 respondents (64.2\%), ADBIS sessions were the most attended. Next are TPDL sessions with 45 of 95 respondents (47.4\%). Yet, this does not mean than ADBIS was more attractive than TPDL, as these figures are actually proportional to the number of sessions for each international conference (5 for ADBIS vs. 3 for TPDL; see also Table~\ref{tab:attendance-sessions} for actual attendance). 

\subsection{Participant Experience}

\begin{figure}[H]
    \centering
    \captionsetup{justification=centering,margin=0cm}
\begin{minipage}{.5\textwidth}
\includegraphics[width=7cm]{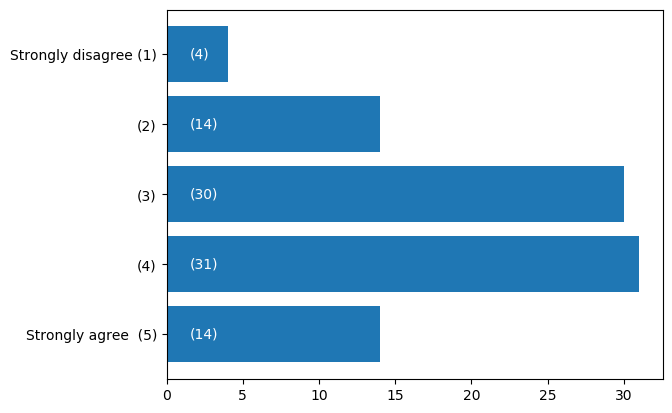} 
\caption{Was the videoconferencing infrastructure adequate for supporting the conferences? (93 responses)}
\label{fig:QB5}
\end{minipage}% This must go next to `\end{minipage}`
\begin{minipage}{.5\textwidth}
\centering
\includegraphics[width=7cm]{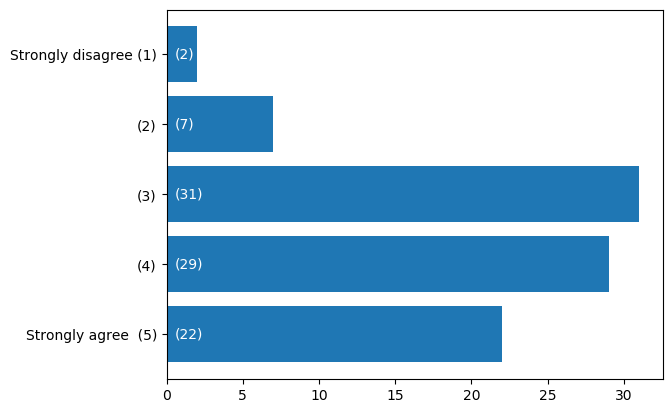} 
\caption{Did the conferences need more social interactions? (91 responses)}
\label{fig:QB6}
\end{minipage}
\end{figure}

\paragraph{(6) The adopted videoconferencing infrastructure was quite acceptable.}
A relative majority of respondents thought the videoconferencing infrastructure was adequate to support all conferences' sessions (Figure~\ref{fig:QB5}).  However, we also note that a significant part of respondents was not fully satisfied. Reasons are mixed: some were related to the participant's configuration, i.e., proxies for firewalls not allowing to connect to our platform (especially on the first day); while others were caused by BBB's limitations (e.g., video broadcasting works best with Firefox and may be limited with other browsers), as well as misconfigurations that had to be fixed almost on the fly by an engineer from the University of Lyon~2, who was a much welcome resource in the organizing committee. 

\begin{figure}[H]
    \centering
    \captionsetup{justification=centering,margin=0cm}
\begin{minipage}{.5\textwidth}
\includegraphics[width=6.5cm]{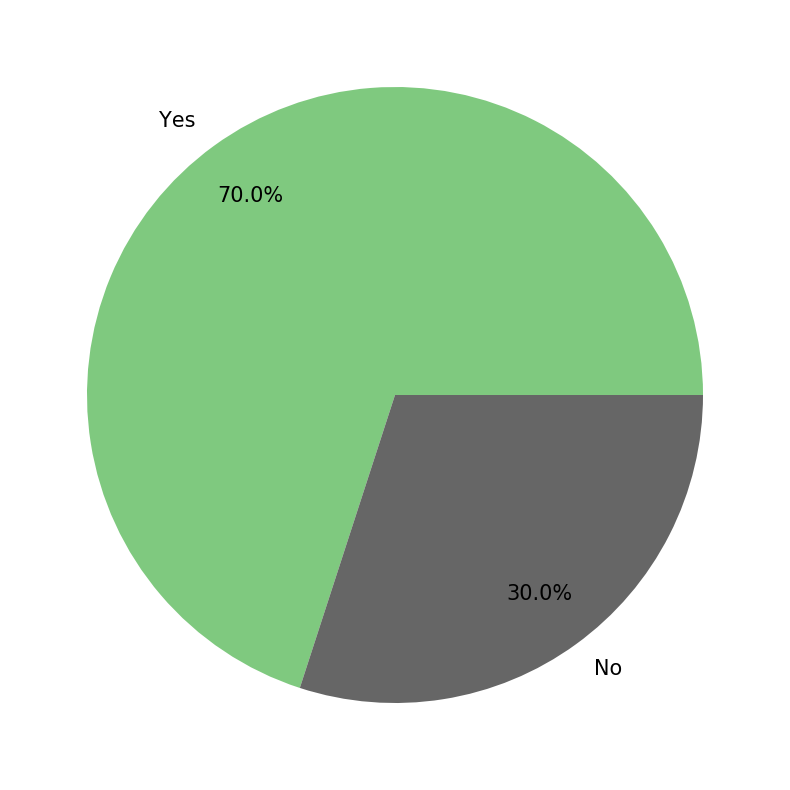} 
\vspace{-0.2cm}
\caption{Did you like the virtual reception? \\(30 responses)}
\label{fig:QB7}
\end{minipage}% This must go next to `\end{minipage}`
\begin{minipage}{.5\textwidth}
\centering
\includegraphics[width=8.5cm]{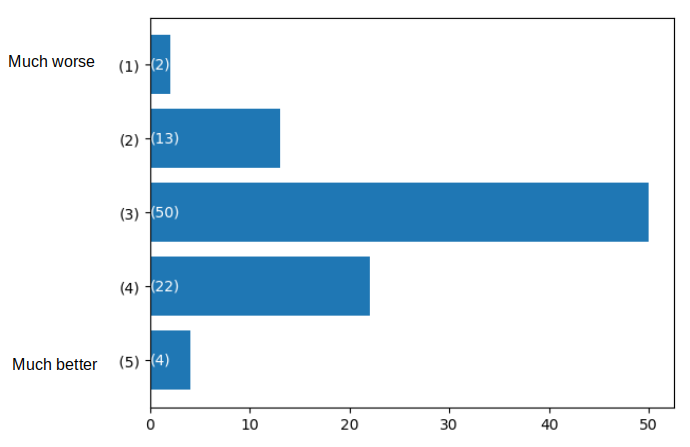} 
\vspace{0.5cm}
\caption{Were ADBIS, TPDL and EDA 2020 better or worse than a typical edition? (91 responses)}
\label{fig:QB8}
\end{minipage}
\end{figure}

\paragraph{(7) There was a lack of social interactions during the conferences.}
As expected, attendants widely agreed that the virtual nature of the conferences induced a lack of social interactions (Figure~\ref{fig:QB6}). 
We did anticipate this issue by proposing a virtual reception session. Although it was reached by few attendants (30/106 respondents), presumably due to the late time (17:45-19:00 CET), we note that 70\% of the participants found it useful (Figure~\ref{fig:QB7}). 
%In addition, these satellite event, especially the  doctoral consortium give a great opportunity to young researchers to discuss on their work. 
Moreover, one of the participants told us that the virtual social session allowed him to realize the changes to be made in his thesis work.
This illustrates a need for more meeting/discussion time, as was highlighted in several comments. We introduced a virtual coffee room on the last day of the conferences, but it was little attended.
%This could therefore be a way amongst other to bring more interactions in future virtual editions.

\paragraph{(8) ADBIS, TPDL and EDA 2020 were neither better nor worse than a typical edition.}
Attendants globally thought the conferences' organization were neither better nor worse than previous editions (Figure~\ref{fig:QB8}). This could be interpreted as a positive appreciation, since it was twice challenging to organize \textit{joint} conferences during a worldwide \textit{crisis} context.

\begin{figure}[H]
    \centering
    \captionsetup{justification=centering,margin=0cm}
\begin{minipage}{.5\textwidth}
\includegraphics[width=8cm]{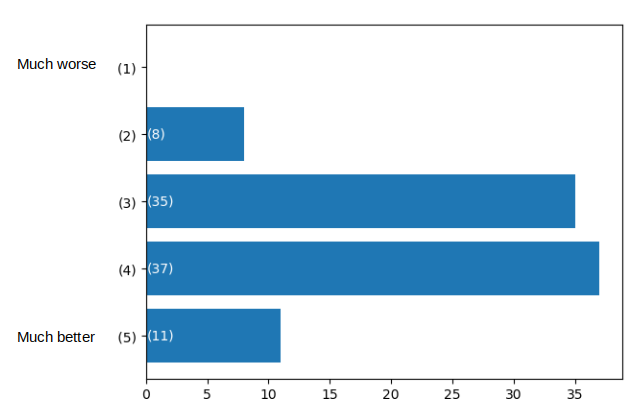} 
\caption{Were ADBIS, TPDL and EDA 2020 better \\or worse than what you expected a virtual conference would be? (91 responses)}
\label{fig:QB9}
\end{minipage}% This must go next to `\end{minipage}`
\begin{minipage}{.5\textwidth}
\centering
\includegraphics[width=8cm]{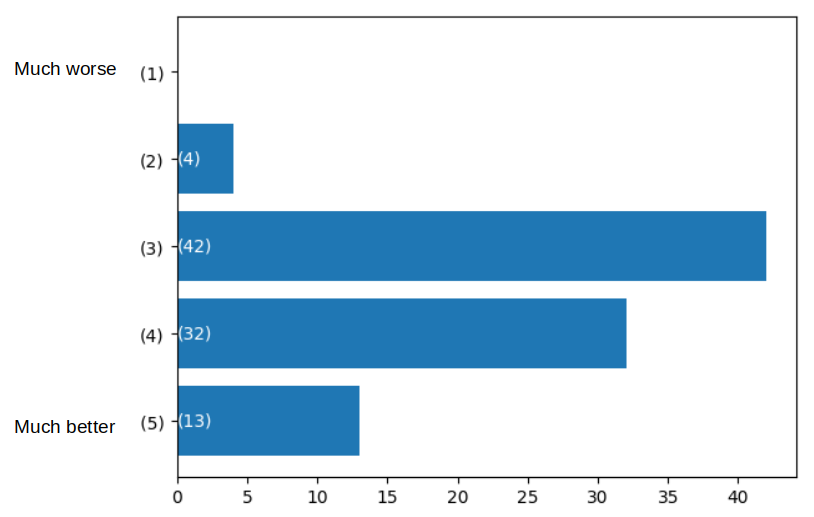} 
\caption{Were ADBIS, TPDL and EDA 2020 better \\ or worse than what you expected from three \\joint conferences? (91 responses)}
\label{fig:QB10}
\end{minipage}
\end{figure}

\paragraph{(9) ADBIS, TPDL and EDA 2020 were better than what participants expected.}
Respondents mostly thought ADBIS, TPDL and EDA 2020 were better than they expected (Figures~\ref{fig:QB9} and \ref{fig:QB10}), confirming that we quite took up the double challenge of organizing a virtual joint conference.

\subsection{Participant Preferences for Future Editions}

\begin{figure}[H]
    \centering
    \captionsetup{justification=centering,margin=0cm}
\begin{minipage}{.5\textwidth}
\vspace{-0.3cm}
\includegraphics[width=7.5cm]{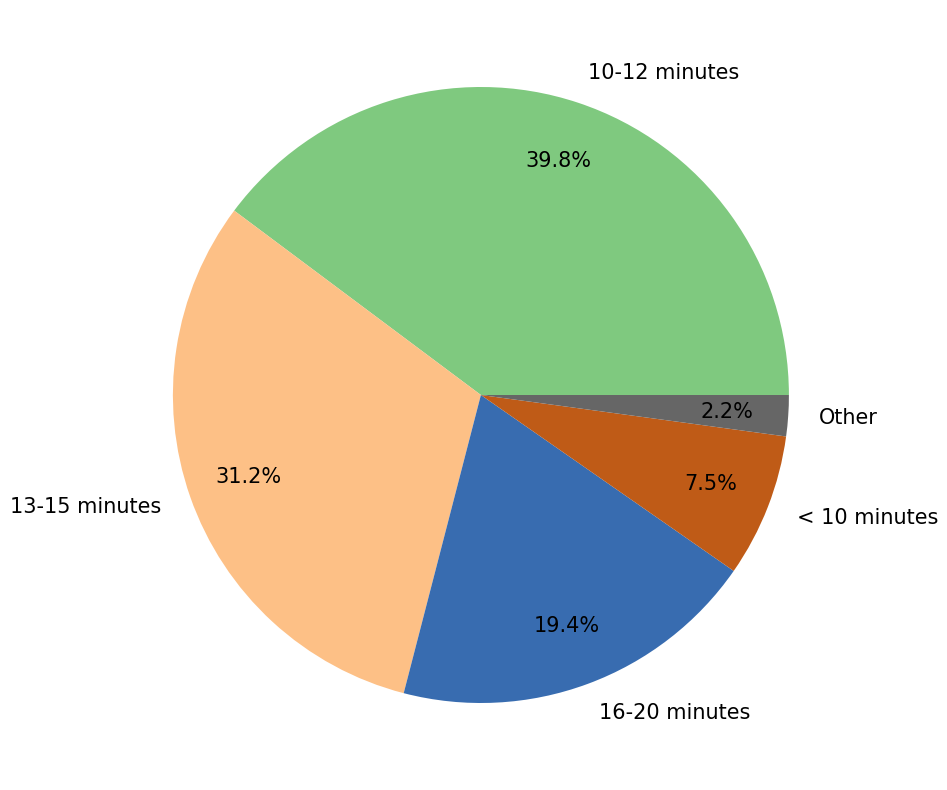} 
\caption{What is the ideal length of a research talk\\ at an online conference? (93 responses)}
\label{fig:QC1}
\end{minipage}% This must go next to `\end{minipage}`
\begin{minipage}{.5\textwidth}
\includegraphics[width=7.5cm]{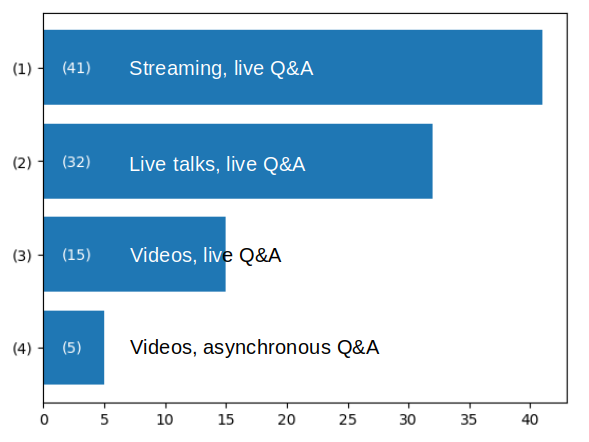} 
\caption{As an attendant, what kind of model do you prefer for research talks? (93 responses)}
\label{fig:QC2}
\end{minipage}% This must go next to `\end{minipage}`
\end{figure}

\paragraph{(10) Attendants have no clear preference on online research talk length.}
The survey did not show any widely preferred length for online research talks. However, we note that the most shared opinion (39.8\% of the respondents) is that talks should take between 10 and 12 minutes (Figure~\ref{fig:QC1}).

\paragraph{(11) Participants prefer talks broadcast in streaming, live Q\&A sessions and having the videos available after the conference.}
58.2\% of the respondents thought the ideal format for research talks was streaming, live Q\&A and having the videos available after the conference (option 1 in Figure~\ref{fig:QC2}), which confirms our choice. 
%This confirms we have chosen the better option for authors talks in ADBIS, TPDL \& EDA 2020. 
Other options were:
\begin{enumerate}
    \setcounter{enumi}{1}
    \item live talks, live Q\&A session and having recorded talks available after the conference;
    \item videos of the talks available beforehand and live Q\&A session;
    \item videos of the talks available beforehand and asynchronous Q\&A session. 
\end{enumerate}

\subsection{Virtualization}

\begin{figure}[H]
    %\centering
    \captionsetup{justification=centering,margin=0cm}
    \begin{minipage}{.5\textwidth}
\centering
\includegraphics[width=8.5cm]{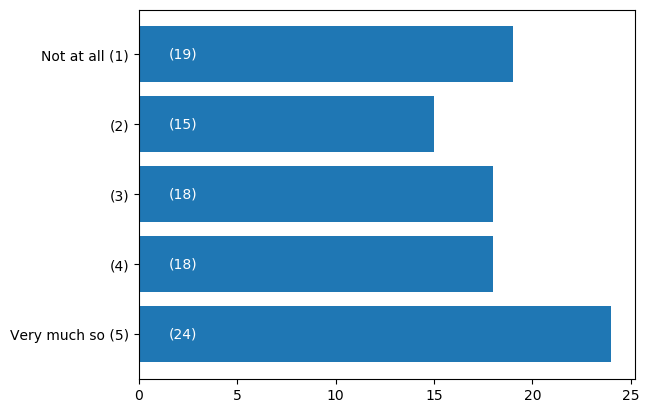}
%\vspace{0.9cm}
\caption{Would you have attended more sessions \\if the conference had been physical? (94 responses)}
\label{fig:QB3}
\end{minipage}
\begin{minipage}{.5\textwidth}
%\centering
\includegraphics[width=8cm]{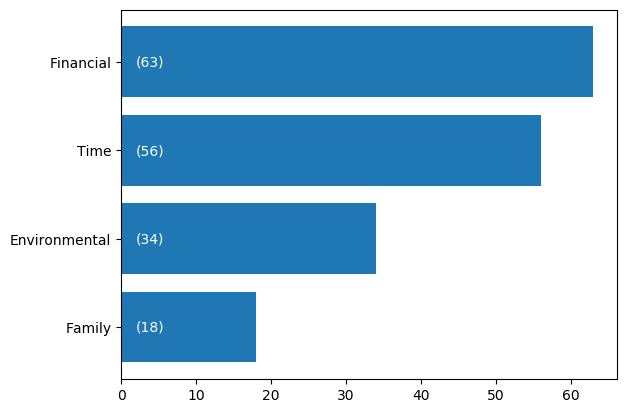} 
\caption{For what reasons would you consider to attend a hybrid conference virtually? (91 responses)}
\label{fig:QC7}
\end{minipage}
\end{figure}

\paragraph{(12) Attendants have no clear preference between virtual and physical sessions.}
We obtained no conclusive result on whether respondents would attend more sessions if the conference was held physically (Figure~\ref{fig:QB3}). 
This could be explained by the fact that the disadvantages of virtual sessions, e.g., distractions or technical issues, are balanced by advantages such as easy switching among sessions. 

\paragraph{(13) Participants thought that the benefits of attending conferences virtually are mainly ``saving money'' and ``saving time''.}
63 out of 91 respondents (69.2\%) estimate that saving money could make them attend a conference virtually if applicable (Figure~\ref{fig:QC7}). Another well shared benefit of virtual conference participation is time saving (61.5\% of the respondents).

\begin{figure}[H]
    %\centering
    \captionsetup{justification=centering,margin=0cm}
\begin{minipage}{.5\textwidth}
\includegraphics[width=8cm]{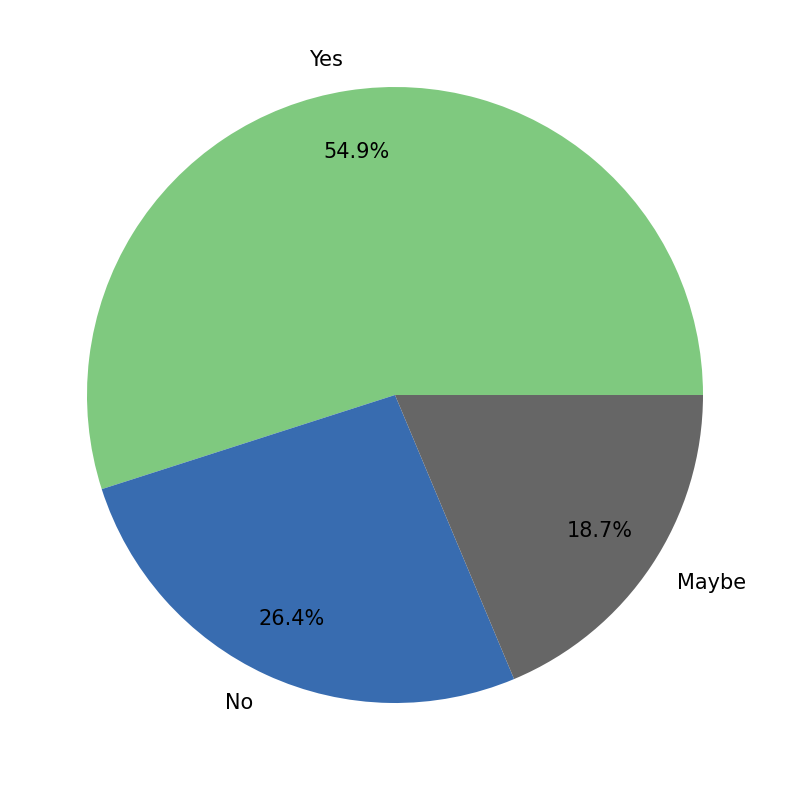} 
\caption{Would you support the idea of alternating physical meetings with purely virtual conferences for reducing CO2 emissions? (91 responses)}
\label{fig:QC9}
\end{minipage}    
\begin{minipage}{.5\textwidth}
\includegraphics[width=8cm]{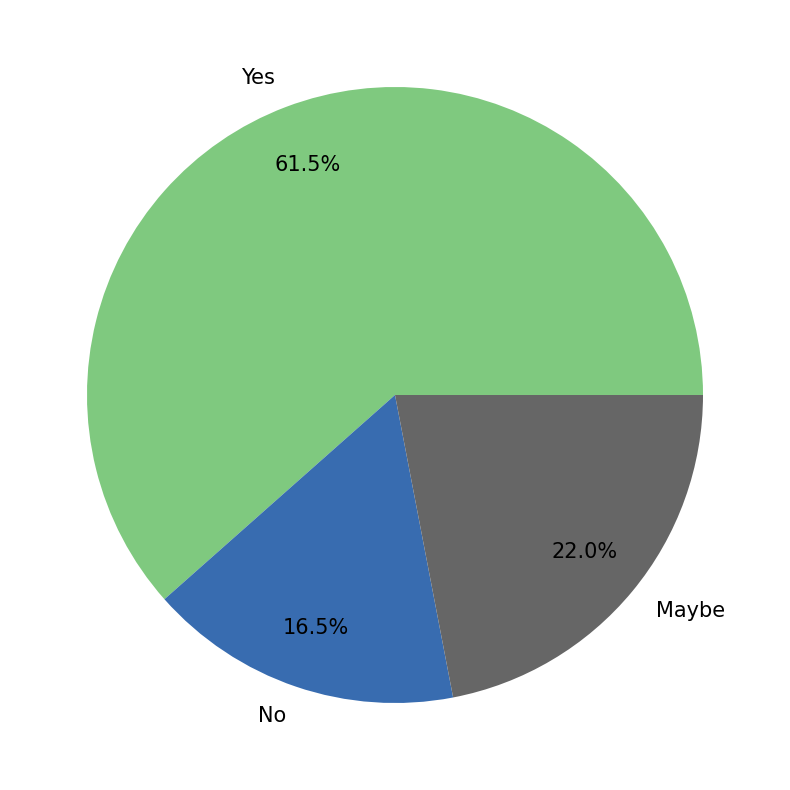} 
\caption{Would you support the idea of having \\hybrid conferences for reducing CO2 emissions? (91 responses)}
\label{fig:QC8}
\end{minipage}% This must go next to `\end{minipage}`
\end{figure}

\paragraph{(14) Participants agree with the idea of alternating physical and virtual conferences to reduce CO2 emissions.}
With climate change increasing, reducing CO2 emissions is an important issue that should be considered when organizing conferences. A first solution for reducing the environmental impact of conferences is to alternate between physical and virtual editions. 54.9\% of the respondents support this idea (Figure~\ref{fig:QC9}).

\paragraph{(15) Participants agree with the idea of hybrid conferences to reduce CO2 emissions.}
Another solution for reducing the environmental impact of  conferences is hybrid organization, i.e., participants can attend virtually or physically, as they prefer. This solution seems the most shared by the respondents (61.5\%; Figure~\ref{fig:QC8}).

\section{Conclusion}
In short, we note that ADBIS, TDPL and EDA 2020 were successfully organized according to the attendants. The organizing committee was able to remedy most of the challenges that arose, although some choices had to be made quite in a hurry and were not fully satisfactory.
We hope that the lessons learned from this first virtual and joint organization will be useful for future editions.

%\section*{Appendix}

\bibliographystyle{apalike} 
\bibliography{feedback-adbis-tpdl-eda-2020} 

\end{document}